\documentclass[twocolumn,showpacs,preprintnumbers,amsmath,amssymb]{revtex4}
\usepackage{graphicx}
\usepackage{dcolumn}
\usepackage{bm}

\begin{document}

\title{Ising model on the Apollonian network with node dependent
interactions}
\author{R. F. S. Andrade$^{1,2}$, J. S. Andrade Jr.$^{2,3}$,
H. J. Herrmann$^{2,3}$}
\affiliation{$^1$Instituto de F\'{i}sica,
Universidade Federal da Bahia,
40210-210, Salvador, Brazil.\\
$^{2}$Computational Physics, IfB, ETH-H\"{o}nggerberg, Schafmattstr. 6,
8093, Z\"{u}rich, Switzerland.
\\$^{3}$Departamento de F\'{i}sica, Universidade Federal do Cear\'{a},
Campus do Pici, 60455-760, Fortaleza, Brazil.}

\date{\today}

\begin{abstract}

This work considers an Ising model on the Apollonian network, where the
exchange constant $J_{i,j}\sim1/(k_ik_j)^\mu$ between two neighboring
spins $(i,j)$ is a function of the degree $k$ of both spins. Using the
exact geometrical construction rule for the network, the thermodynamical
and magnetic properties are evaluated by iterating a system of discrete
maps that allows for very precise results in the thermodynamic limit. The
results can be compared to the predictions of a general framework for
spins models on scale-free networks, where the node distribution $P(k)\sim
k^{-\gamma}$, with node dependent interacting constants. We observe that,
by increasing $\mu$, the critical behavior of the model changes, from a
phase transition at $T=\infty$ for a uniform system $(\mu=0)$, to a $T=0$
phase transition when $\mu=1$: in the thermodynamic limit, the system
shows no exactly critical behavior at a finite temperature. The
magnetization and magnetic susceptibility are found to present
non-critical scaling properties.
\end{abstract}

+

\maketitle

\newpage\

\section{Introduction}

Magnetic models on complex networks have quite distinct behavior from
those on regular lattices \cite{Dorogo2008}. Their properties are of far
greater importance than just a mathematical curiosity, since they
establish landmarks for the behavior of many systems, like social,
economic, and communication networks. For such systems, the understanding
of the conditions leading to a phase transition, or a sudden collective
change in the behavior of the agents, is of utmost importance to avoid a
breakdown of social structures or collective current day technological
facilities \cite{CostaAdPhys,Boccaletti}.

The absence of a finite temperature phase transition in the thermodynamic
limit $T\rightarrow\infty$, where $N$ is the number of nodes
\cite{Dorogo2002,Stauffer02}, concomitant with the presence of a finite
degree of magnetic ordering, stays among the first results that have been
obtained for Ising models on the standard Barabasi-Albert (BA) scale-free
network \cite{Barabasi99}, where the exponent of the node distribution
$P(k)\sim k^{-\gamma}$ assumes the value $\gamma=3$. It was also observed
that finite temperature critical behavior is found when $\gamma \in
(3,5]$, while, for $\gamma>5$ the critical behavior collapses at $T=0$.
Later, an interesting interplay between critical behavior and node
dependent interaction constants has been evidenced
\cite{Indekeu2005,Indekeu2006}: if the strength of interactions in a BA
network, with a given value $\gamma$, is non-uniformly reduced according
to
\begin{equation}\label{eq1}
J_{i,j}=J_0/(k_ik_j)^\mu,
\end{equation}
where $k_\ell$ is the degree of node $\ell$, the critical behavior moves
into the universality class of the uniform model with a different value
$\gamma ^{'}$. This makes it possible, for instance, to devise models in
the standard BA network that undergo finite temperature phase transition.
An analytic expression
\begin{equation}\label{eq2}
\gamma'=(\gamma-\mu)/(1-\mu)
\end{equation}
has been derived based on scaling arguments but, although it has been
numerically verified for BA networks, it is not known whether its validity
extends to other networks.

The purpose of this work is to investigate the effect of a node dependent
coupling constant on the properties of an Ising model on the Apollonian
network (AN) \cite{Andrade04,Doye04}. This network has very special
features \cite{Soares,Moreira}, including presenting a power law
distribution of node degrees, with exponent $\gamma\simeq 2.58$. Previous
results for Ising models on the AN have shown that, for a variety of
situations where both ferro- and antiferromagnetic interactions are
allowed, phase transition in the thermodynamic limit occur only at
$T=\infty$ \cite{Andrade04,RAndrade05}. AN's are constructed according to
precise geometrical rules, which lead to exact self similar patterns and
scaling properties. They are also amenable to mathematical analysis based
on renormalization or inflation methods, as the transfer matrix (TM)
formalism we will use here, which allow for the evaluation of its
properties in the thermodynamic limit.

These facts turn this model particularly suited for testing the existence
of a finite temperature phase by modulating the coupling constants
according to Eq. (\ref{eq1}). On the other hand, since the AN geometric
rules lead to a well defined value of $\gamma$, there is no general free parameter
we can use in the study to verify the validity of Eq. (\ref{eq2}).
Further, it must be stressed that, despite the fact that AN displays power
law distribution of node degree, it differs substantially from BA network
with respect to other topological properties, as the existence of many
closed loops. This is expressed, among other measures, by the clustering
coefficient $C$, which is very high $(\sim 0.85)$ for AN and very small
$(\sim 1/N)$ for the BA \cite{Andrade04,Doye04}.

The rest of this paper is organized as follows: Section 2 introduces the
basic properties of AN networks and of the proposed model; details of the
used TM scheme to evaluate the thermodynamical properties are discussed in
Section 3. We discuss our main results in Section 4, emphasizing the
emergence of a cross-over in  $\mu$. Finally, Section 5 closes the
paper with our concluding remarks.

\section{Apollonian network and model}

AN's have been recently introduced in the complex network framework
\cite{Andrade04,Doye04}, although the original concepts can be traced back
to ancient Greece, where the problem of optimally filling two and three
dimensional spaces with circles and spheres has been studied by Apollonius
of Perga \cite{Herrmann90}. The complex solution to this problem, which
amounts to placing tangent structures with well defined radii at precise
centers, suggests the far simpler problem of constructing the AN. In this
case, one just has to put a node in each circle center, and a network edge
between the centers of each pair of tangent circles. This process can be
followed in a recursive way in terms of the generation $g$ in which new
circles are added to the structure.  In this work, we consider that, at
the zeroth generation $g=0$, three tangent circles with the same radius
occupy the centers of an equilateral triangle (see Fig. 1). For the
$g+1$-th generation, the network construction consists in putting a node
within each triangle of the $g$-th generation, and connecting it to each
of the triangle corners. It is a simple matter to verify that the number
of network nodes $N(g)$ and edges $B(g)$ increase according to,
respectively, $N(g)=(3^{g}+5)/2$ and $B(g)=(3^{g+1}+3)/2$. The average
number of neighbors per node equals 6, since $B(g)/N(g)\rightarrow 3$ in
the limit $g\rightarrow\infty$.

\begin{figure}
\begin{center}
\includegraphics*[width=4.2cm,height=3.cm,angle=0]{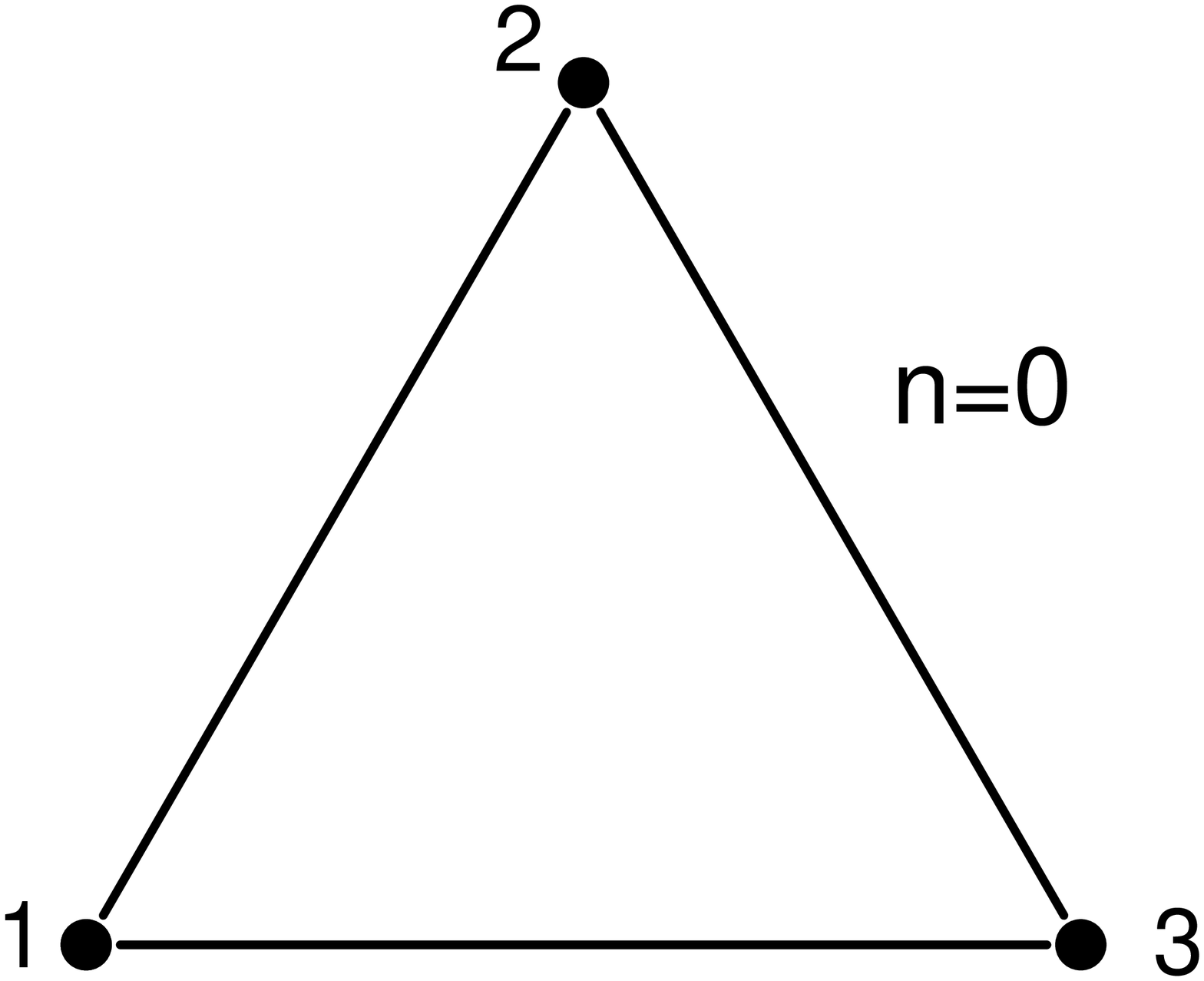}
\includegraphics*[width=4.2cm,height=3.cm,angle=0]{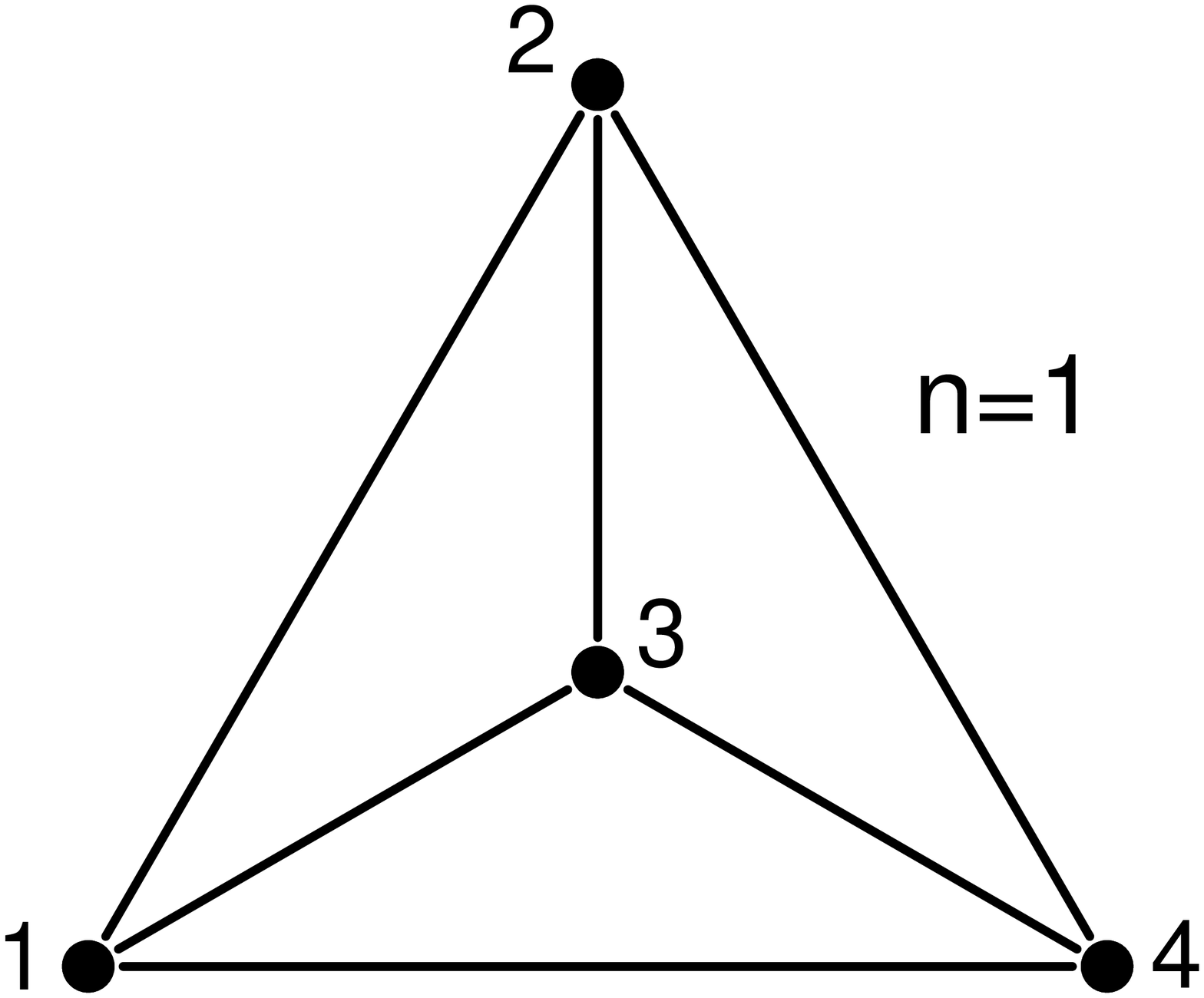}
\includegraphics*[width=4.2cm,height=3.cm,angle=0]{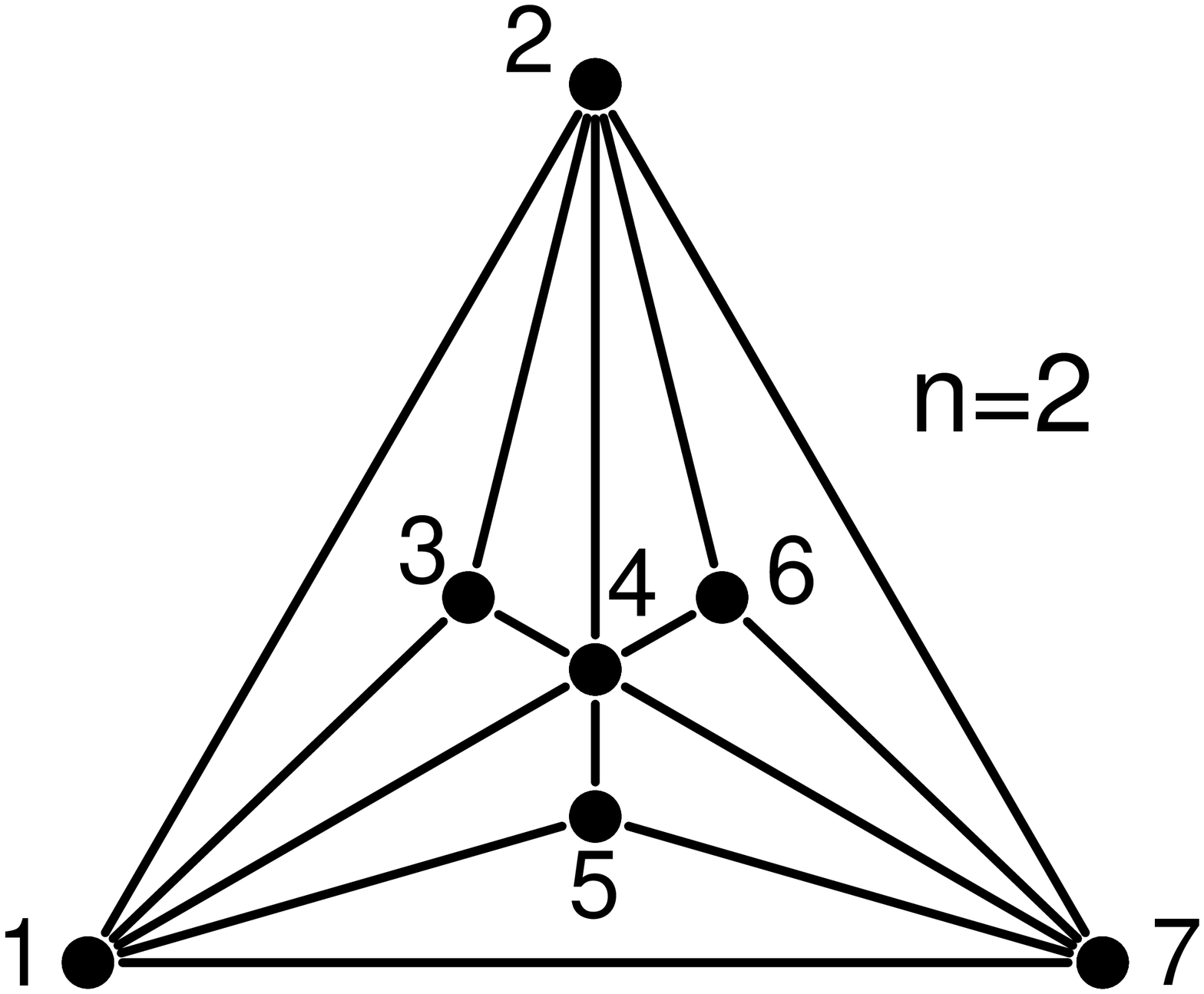}
\end{center}
\caption{Geometrical construction of the first three generations ($g=0,1,$
and $2$) of the AN. Nodes are numbered according to the scheme used in
\cite{RAndrade05}} \label{fig1}
\end{figure}

For a given generation $g>0$, the largest node degree is $k_c(g)=3\times
2^{g-1}$, where the subscript $c$ indicates that such node occupies the
central network position. The second largest degree nodes, with
$k_e(g)=2^g+1$, occupy the external corners. At any generation $g$, there
will be nodes with degree $k=k_c(\overline{g}),\overline{g}=1,\ldots , g$
and $k_e$. The degree dependent node multiplicity is
$m(\overline{g})=3^{g-\overline{g}}$ for the internal nodes, and $m(g)=3$
$\forall g$, for the nodes at outer network corners.

As already quoted, the resulting AN is scale free. However, it also has
other properties that are typical for other complex network classes, as
being small world (mean minimal path $\langle \ell \rangle \sim \ln N$),
hierarchical (the clustering coefficient of individual nodes $c(k)$ has a
power law dependence on $k$), and having a large clustering coefficient
$C$. Because of this, systematic network clustering analysis based on
several independent measures \cite{CostaAndrade} shows that AN does not
belong to the same class as the most studied network sets, generated by
the algorithms proposed by Watts and Strogatz \cite{Watts98} and Barabasi
and Albert \cite{Barabasi99}.

We consider the Ising model with spins $\sigma_i=\pm1$  placed on each
site of the Apollonian network. Pairs of spins $(i,j)$, which are
neighbors on the network, interact with coupling constants $J_{i,j}$.
Thus, the Hamiltonian for the system can be written as
\begin{equation}\label{eq3}
H_g = -\sum_{(i,j)}J_{i,j}\sigma_i \sigma_j - h \sum_{i} \sigma_i,
\end{equation}
where $J_{i,j}$ is given by Eq.(\ref{eq1}). In our previous studies, we have
considered inhomogeneous models, in which the constants $J_{i,j}$ depend
on the generation $g$ at which the edge, hence the second spin in the
pair, was introduced into the network. Due to the fact that, at each
generation, the newly introduced nodes are connected to nodes that were
introduced in previous generations, the scheme introduced in Ref.
\cite{RAndrade05} does not assign the values of $J_{i,j}$ according to the
rule of Eq. (\ref{eq1}). In the following Section we discuss how to
implement  the interaction constants of Eq. (\ref{eq1}) in connection with
the TM method used to evaluate the model properties.

\section{TM recurrence maps}

The basic steps to implement the TM method we use to evaluate the
thermodynamic properties have been presented, with some detail, in one of
our previous works \cite{RAndrade05}. However, the method needs to be
adapted to the specific situation introduced by the more complex
interaction given in Eq. (\ref{eq1}). Thus, let us briefly recall that the
TM scheme amounts to write down the partition function $Z(T,h,N(g))$,
where $H$ denotes the magnetic field, for any value of $g$ in terms of a
TM that describes the interactions between any two of the outer AN sites.
In this process, it is necessary to perform a partial trace over all
interaction dependent configurations. Due to the exact geometric AN
construction rule, it is possible to express the TM matrix elements at
generation $g+1$ in terms of the corresponding elements at generation $g$.
In this framework, we basically work with a set of $2 \times 2$ square
matrices
\begin{equation}\label{eq4}
M_g=
\begin{pmatrix}
 a_{g} & b_{g} \\
 c_{g} & d_{g}
\end{pmatrix},
\end{equation}
and a set of non-square auxiliary matrices
\begin{equation}\label{eq5}
L_g=
\begin{pmatrix}
 p_{g} & q_{g} & r_{g} & s_{g} \\
 t_{g} & u_{g} & v_{g} & w_{g}
\end{pmatrix},
\end{equation}
which explicitly include the dependence of the third outer node spin
variable. As the $L_g$ matrix elements are numbered according to the
lexicographic order, the following relations hold:
$a_{g}=p_{g}+q_{g}$, $b_{g}=r_{g}+s_{g}$, $c_{g}=t_{g}+u_{g}$,
$d_{g}=v_{g}+w_{g}$. For more symmetrical models, and field independent
situations, the number of independent variables can be reduced.

For the homogeneous systems, it was possible to write down a single set of
recurrence relations between matrix elements in successive generations.
Although the basic idea of the method remains the same, for the current
model, it is necessary to track the way the nodes are reconnected when
they go from $g$ to $g+1$. This influences the change in their degrees, so
that the same node will contribute differently for distinct values of $g$.

We start the discussion of the changes in the TM scheme by pointing
out that, besides knowing the set of node degrees and corresponding degree
(node)  multiplicity, it is necessary to go one step further, and identify
each of the $P(g)=(g^2-g+2)/2$ different triangles in which the $g$
network can be disassembled. In this respect, each triangle is
characterized by the node degrees $k_i,k_j,$ and $k_\ell$ of the nodes
$i,j,$ and $\ell$, respectively. For any triangle and any $g>1$, there is
always (only) one node with $k=3$. Note that $P(g)$ grows only with the
square of $g$, so that, even for a complex interaction structure, there is
practically no constraint to numerically compute these matrix elements
for very large values of $g$.

Once this set has been identified, we evaluate the model properties at
generation $g$ by computing the contribution to the partition function
from each of these $P(g)$ triangles, storing them in corresponding TM's
$L_{g}^i$, $i=1, \ldots, P(g)$.

To proceed further, we must consider that the $g$
evaluation is equivalent to the $g-1$ one, provided we start
with triangular units with a fourth node added at the central position.
This way, it is possible to compute the contribution of the new
$P(g-1)$ triangles, by performing partial trace over the
contributions from the central node of each of these structures. The new
form of the general recurrence relations for the matrix elements,
\begin{equation}\label{eq6}
(L_{g+1})_{i,k}^\alpha = \sum_{\ell}(L_g)_{i,j\ell}^\eta (L_g)_{i,\ell
k}^\epsilon (L_g^t)_{k,j\ell}^\delta,
\end{equation}
is quite similar to that of the uniform model. The difference refers to
the superscripts $\eta, \epsilon,$ and $\kappa$, which identify which
three TM's (corresponding triangles) have been put together. The same
arguments can be used again, until we obtain one single TM that accounts
for the contributions of all network nodes.

The results we present in the next Section consider $g\leq 50$ which,
for the largest value, is roughly of the order of magnitude of the Avogrado
number. The adaptation of the uniform TM procedure to
take into account the node dependent interaction constant depends
basically in the identification of the basic triangular units and the
assembling rules that combine them when one goes from $g$ to $g+1$. A
summary of the implementation of the details is provided in the Appendix.

Finally, it is important to note that the map iterations can be more
conveniently performed if we rewrite the set of recurrence maps given by
Eq. (\ref{eq6}) in terms of the free energy and the ratio of the $L_g$
matrix elements to the largest one $(q_g)$. Indeed, this avoids numerical
divergences, in the low temperature region, when $g$ increases, as
conveniently discussed in Ref. \cite{RAndrade05}.

\begin{figure}
\begin{center}
\includegraphics*[width=8cm,height=5.5cm,angle=0]{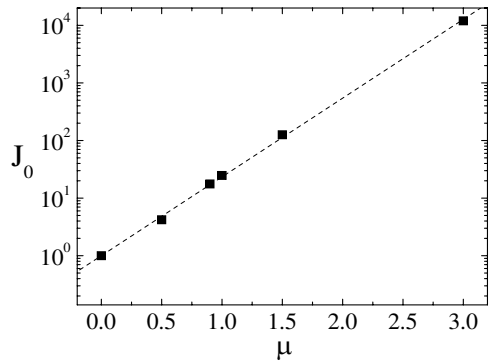}
\end{center}
\caption{Dependence of $J_0$ on $\mu$, as a result from fixing the
ground state energy per spin $u_0=-3$.} \label{fig2}
\end{figure}

\section {Results}

According to the previous Section, we present results for fixed number $g$
of generations, usually  $g=10, 20, 30, 40,$ and $50$, corresponding to
networks with $2.953\times10^4, 1.743\times10^9, 1.030\times10^{14},
6.079\times10^{19}$ and $3.590\times10^{23}$ sites, respectively. The
precise numerical evaluation of the free energy $f(T,h)$ allows to obtain
the entropy $s(T)$, specific heat $c(T)$, magnetization $m(T,h=0)$, and
susceptibility $\chi(T,h=0)$. It is also possible to calculate the ratio
$\lambda_1/\lambda_2$ of the two TM eigenvalues. For models on Euclidian
lattice, as well as on hierarchical and several fractal structures
\cite{Andrade00}, this quantity is directly related to the correlation
length $\xi$. In the case of complex networks and, in particular, of the
AN, the connection between these quantities is not so obvious, as the
distance between two outer nodes remains always 1 in any generation.
Therefore, we will discuss the behavior of $1/ln(\lambda_1/\lambda_2)$,
although we refrain ourselves from calling it $\xi$.

\begin{figure}
\begin{center}
\includegraphics*[width=8cm,height=5.5cm,angle=0]{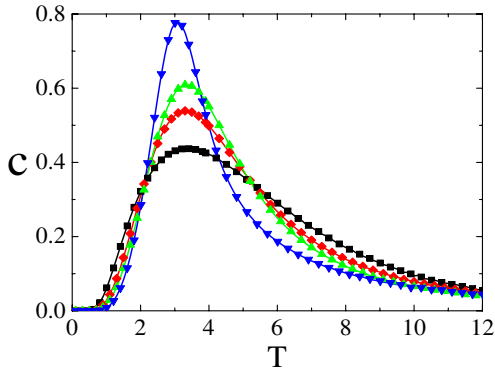}
\end{center}
\caption{Plots of the specific heat $c$ for $g=40$ (symbols) and $g=50$
(solid line) for different values of $\mu$: 0 (squares), 0.5 (diamonds),
1.0 (up-triangles) and 3.0 (down-triangles). The superposition of curves
and symbols for $g=40$ and $g=50$ indicates that $c$ converged to its
value in the thermodynamic limit.} \label{fig3}
\end{figure}

Eq. (\ref{eq1}) indicates that the coupling constants $J_{i,j}$ linearly
depend on $J_0$. According to Section 2, the number of connections in the
AN at generation $g$ is $L_g=(3^{g+1}+3)/2$. If we take $J_0=1$ when
$\mu=0$, the free energy per spin in the $g\rightarrow\infty$ limit is
$u_0=-3$ at $T=0$.  If we fix $J_0=1$ and let $\mu$ increase, the value of
$u_0$ decreases and, besides that, all thermodynamic effects will occur at
a lower value of $T$. Thus, to avoid choosing an adequate temperature
scale to work with at each value of $\mu$, we find it more convenient to
choose a $\mu$ dependent value $J_0(\mu)$, by requiring that $u_0=-3,
\forall \mu$. In Fig. 2, we show the dependence of $J_0(\mu)$ on $\mu$,
which shows that $J_0\sim \exp(\mu)$. As a consequence of this choice, all
maxima of the specific heat occur roughly at the same value of $T$.

Fig. 3 shows, for $g=40$ and $50$ and $\mu=0,0.5,1$, and $3$, that
the curves for $c$ are rather insensitive to the values of $g$ and
$\mu$. Moreover, they are completely smooth, with a Schottky like
maximum at a temperature $T_s$. This constitutes a main difference
to the results for the $\gamma=3$ BA networks \cite{Indekeu2005}.
There is reported the presence of a finite critical temperature,
identified by a jump in the specific heat when $\mu>0.5$, which
changes into having a diverging slope when $0.5> \mu > 0.33$.
According to Eq. (\ref{eq2}) and to the AN known value of
$\gamma\simeq 2.58$, similar critical behaviors should emerge for
$\mu
> 0.61$ and $0.61> \mu > 0.47$, if the AN were to fall within the BA
universality class. This clearly shows that the validity of Eq.
(\ref{eq2}) can not, in general, be extended from the BA to other network
classes, even if they are scale free as the AN.

The same calculations reveal that the resulting patterns for $m,\chi$ and
$1/ln(\lambda_1/\lambda_2)$ depend, first, on the generation $g$, and
further on whether $0<\mu<1$, $\mu=1$ and $\mu>1$. So it is adequate to
discuss them separately.

\subsection {$0<\mu<1$}

Within this parameter interval, we observe that, like for the uniform
model, $1/ln(\lambda_1/\lambda_2)$ numerically diverges $(>10^{100})$ for
a non-zero temperature $T_d(g)$, which increases linearly with $g$. Since
$g\sim \ln N$, $T_d$ depends in a logarithmic way on the system size. If
we write $T_d(g)\sim A(\mu) \ln N$, we find that $A(\mu)$ decreases with
$\mu$ (see Fig. 4).

\begin{figure}
\begin{center}
\includegraphics*[width=7cm,height=5cm,angle=0]{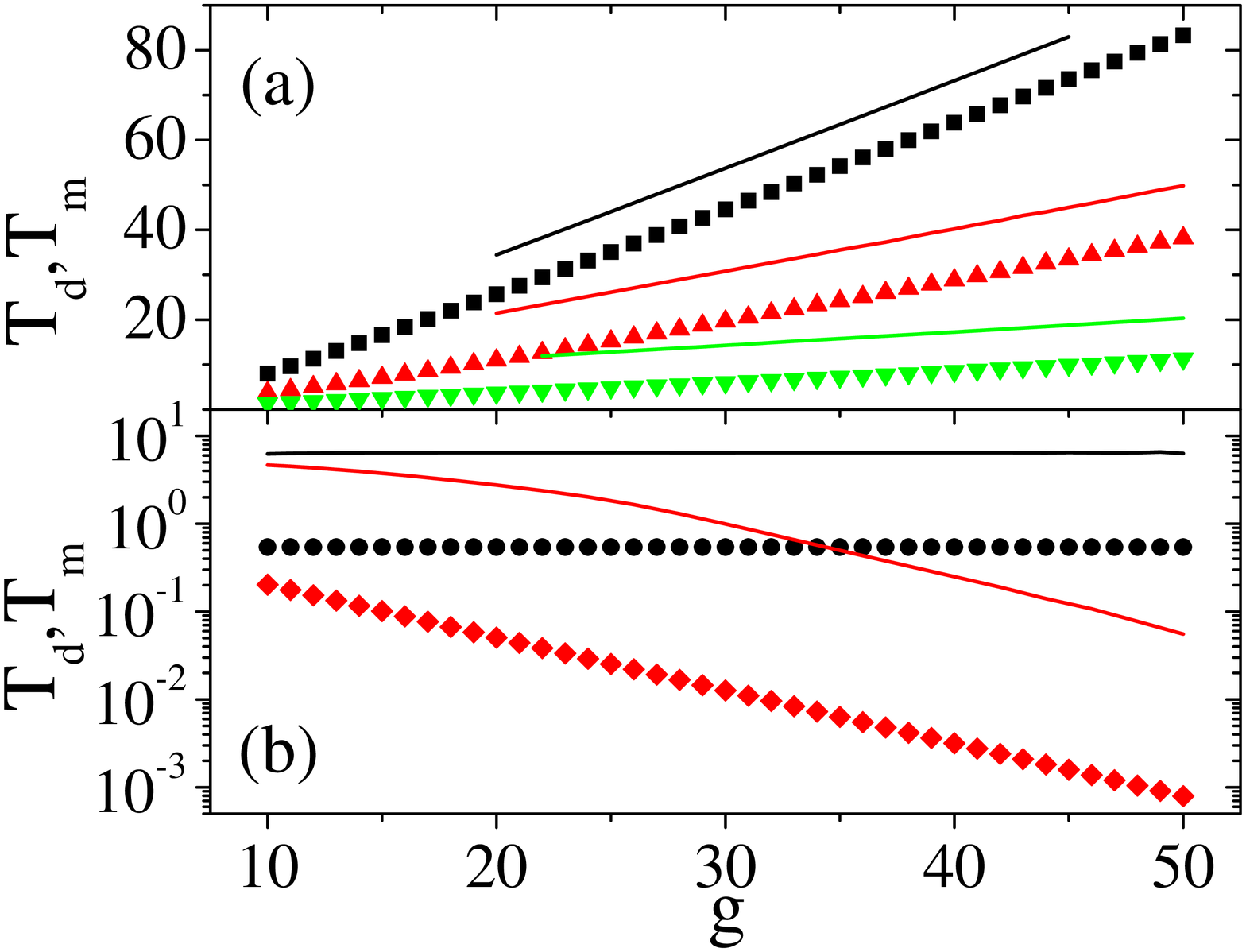}
\end{center}
\caption{Dependence of the temperatures $T_d$ (symbols) and $T_m$ (solid
lines) on the network generation $g$. (a) $\mu=0.0$ (squares), $0.5$
(up-triangles) and $0.8$ (down-triangles). (b) $\mu=1.0$ (circles) and
$1.2$ (diamonds).} \label{fig4}
\end{figure}

In Fig. 5, the behavior of the zero field magnetization $m(T,h=0,N(g))$,
which is exactly one when $T=0$, slowly decreases when $T$ increases. Its
behavior when $\mu=0.5$, at larger values of $T$, is a bit more complex
than that for $\mu=0$ (Fig. 5a). There it is clear that $m$ suffers a
first cross-over to an exponential decay at $T_s$, which is followed by a
transition to a second exponential decay, mediated by a larger constant,
at $T_m(g)$. The magnetization curves for different $g$ collapse during
the first and second regimes. The third regime will later on be
interrupted again by a smoother decay.  As observed with $T_d(g)$,
$T_m(g)\sim B(\mu) g$, with $B(\mu)\sim A(\mu)$. However, $T_d$ and $T_m$
do not coincide. The second part of the magnetization curves, where
$m(T,0,N(g))$ overlaps for different values of $g$, extends over wider $T$
intervals when $g$ increases. This shows that, in the thermodynamical
limit $g\rightarrow\infty$, the value of $m$ will follow the second
exponential decay when $T\rightarrow\infty$. Nevertheless, as observed for
$T_d$, this region grows logarithmically with the network size.

The behavior of $\chi$ is strongly correlated with that of $m$. It
vanishes when $T\rightarrow 0$, then it grows with $T$, shows a first
maximum at a $g$ independent $T_s$, and a second $g$ dependent maximum at
$T_m$.  As for $m$, the curves for larger values of $g$ overlap for much
larger distances. The maxima of the $\chi$ curves are described by an
universal function, as can be observed in the very precise re-scaled
curves in Fig. 6a, which shows that the scaling exponents increase with
$\mu$. Note that only the value of $\chi$ needs to be scaled by the
corresponding maxima, while the location at the temperature axis is
corrected by shifting the scale by $T_m(g)$. This excludes any possibility
of having a critical phenomenon associated with susceptibility maxima.

\subsection {$\mu=1$}

\begin{figure}
\begin{center}
\includegraphics*[width=8cm,height=5.5cm,angle=0]{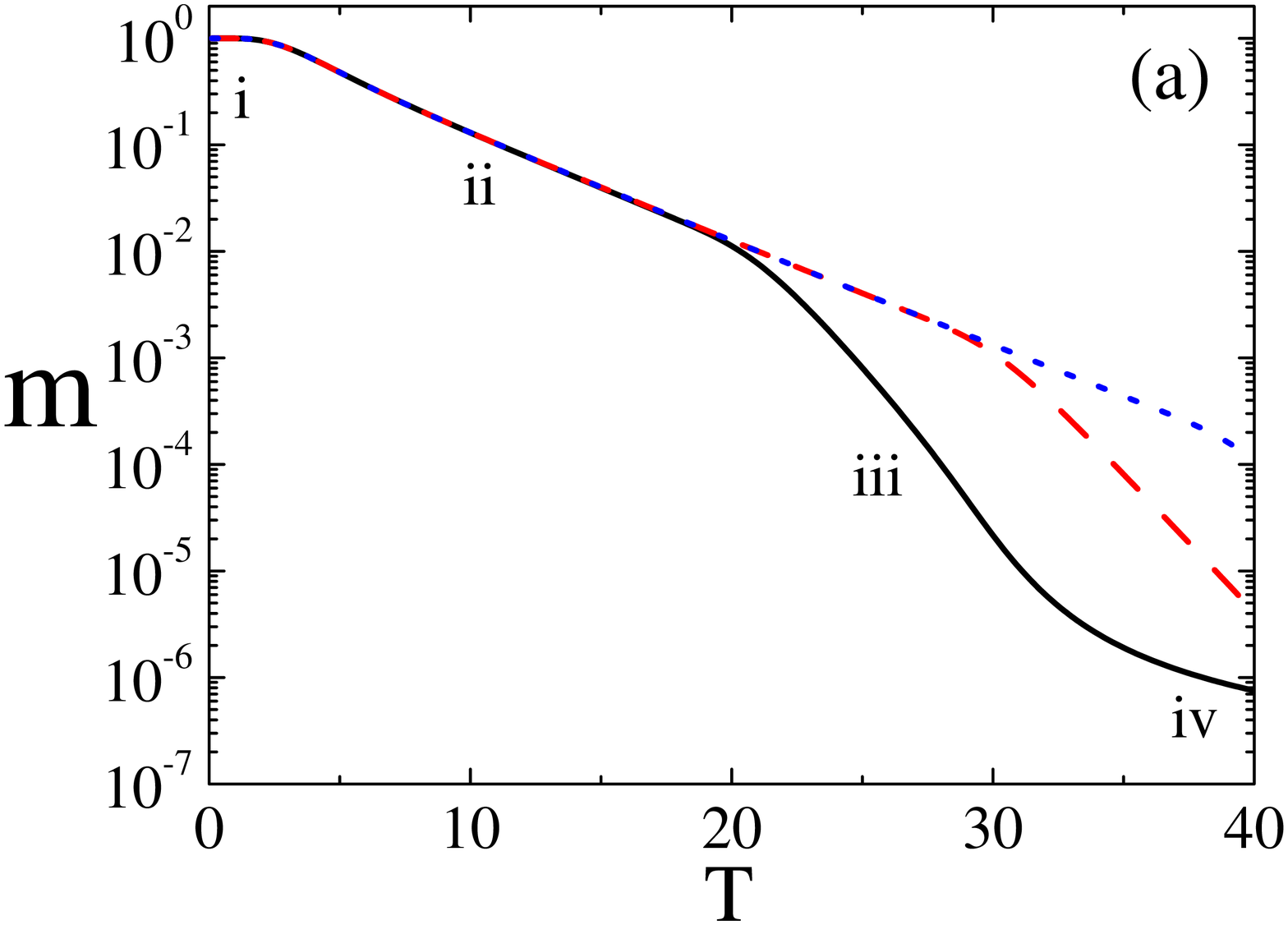}
\includegraphics*[width=8cm,height=5.5cm,angle=0]{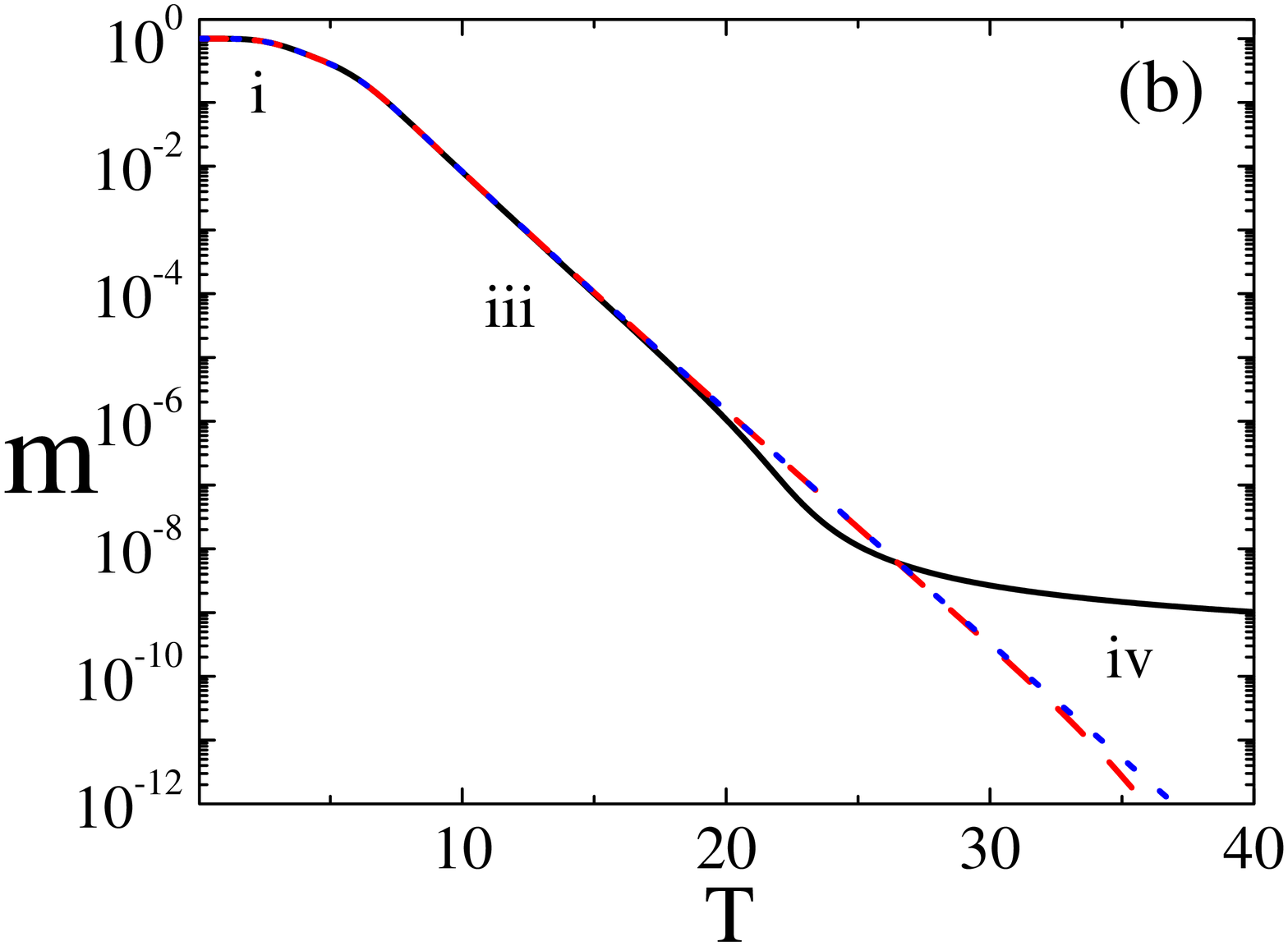}
\includegraphics*[width=8cm,height=5.5cm,angle=0]{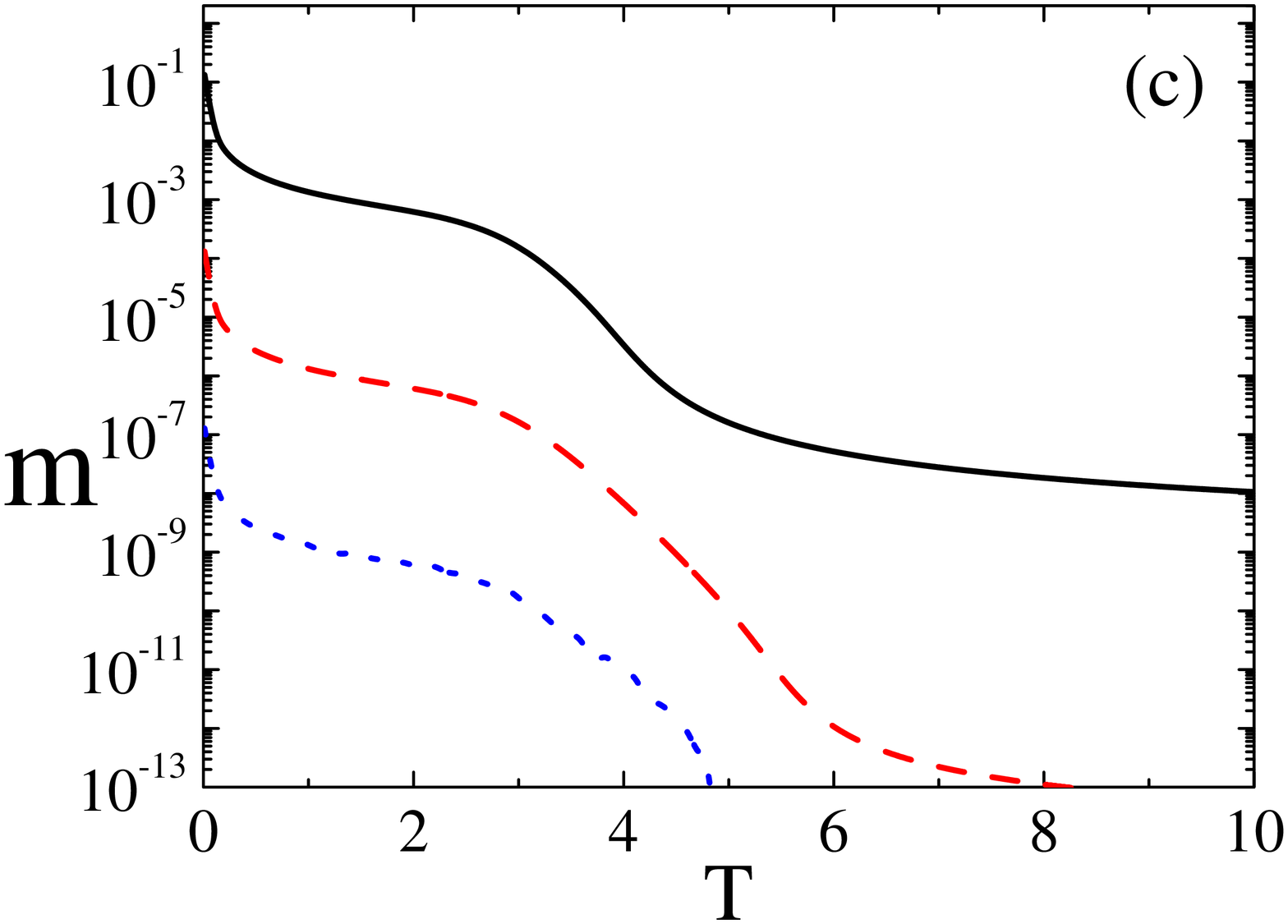}
\end{center}
\caption{Behavior of $m(T,h=0;N(g))$ against $T$ for different values of
$g$ when $\mu=0.5,1.0$ and $1.5$. (a) $\mu=0.5$, $g=20,30$ and $40$
indicated by dots, dashes and solid lines. Four regions characterized by
different behavior are obtained: Two of them are $g$ independent, the
second of which has an exponential decay. The third region starts with a
cross-over to a second exponential regime. (b) $\mu=1.0$, same symbols as
in (a). The first exponential region has disappeared. (c) $\mu=1.5$,
$g=10,15$ and $20$ indicated by dots, dashes and solid lines. The first
region in (a) has disappeared. Two $T$ intervals separated by a $g$
dependent crossover temperature are observed. As $g$ increases, $m$
vanishes for any $T>0$.} \label{fig5}
\end{figure}

This value of $\mu$ determines a crossover in the behavior of the system,
which is reflected both in $m$ and $\chi$. This change can be noticed in
Fig. 3, which shows that $A(\mu=1)=B(\mu=1)=0$, i.e., the temperatures
associated with the maxima of the susceptibility and the divergence of
$1/ln(\lambda_1/\lambda_2)$ become independent of the system size. The
precise value of $T_d$ depends, of course, on the threshold value of the
numerical divergence. However, by plotting the value of
$1/ln(\lambda_1/\lambda_2$ as function of $1/T$, we notice a linear
dependence in the $T\rightarrow 0$ limit, suggesting that $T_d=0$.

However, this new behavior cannot be associated with the emergence
of criticality. First we recall that Fig. 3 does not indicate any
change in the Schottky profile and, second, we see that $T_m>T_s$.
Finally, the $\chi$ curves in the region around $T_m$, which shows a
perfect scaling with respect to $g$ with scaling exponent 1, are
completely smooth (see Fig. 6b). Note that the horizontal axis
indicates that it is not necessary to shift temperature as in Fig.
6a. Note that the two maxima, which were observed when $\mu<1$, have
merged together, and that the large temperature side is
characterized by an exponential decay.

As for the previous $\mu$ interval, the behavior of $m$ is strongly
correlated with that for $\chi$. It is characterized by a single
exponential decay after $T_m$, with a very large constant, as shown in
Fig. 5b.

\subsection {$\mu>1$}

\begin{figure}
\begin{center}
\includegraphics*[width=8cm,height=5.5cm,angle=0]{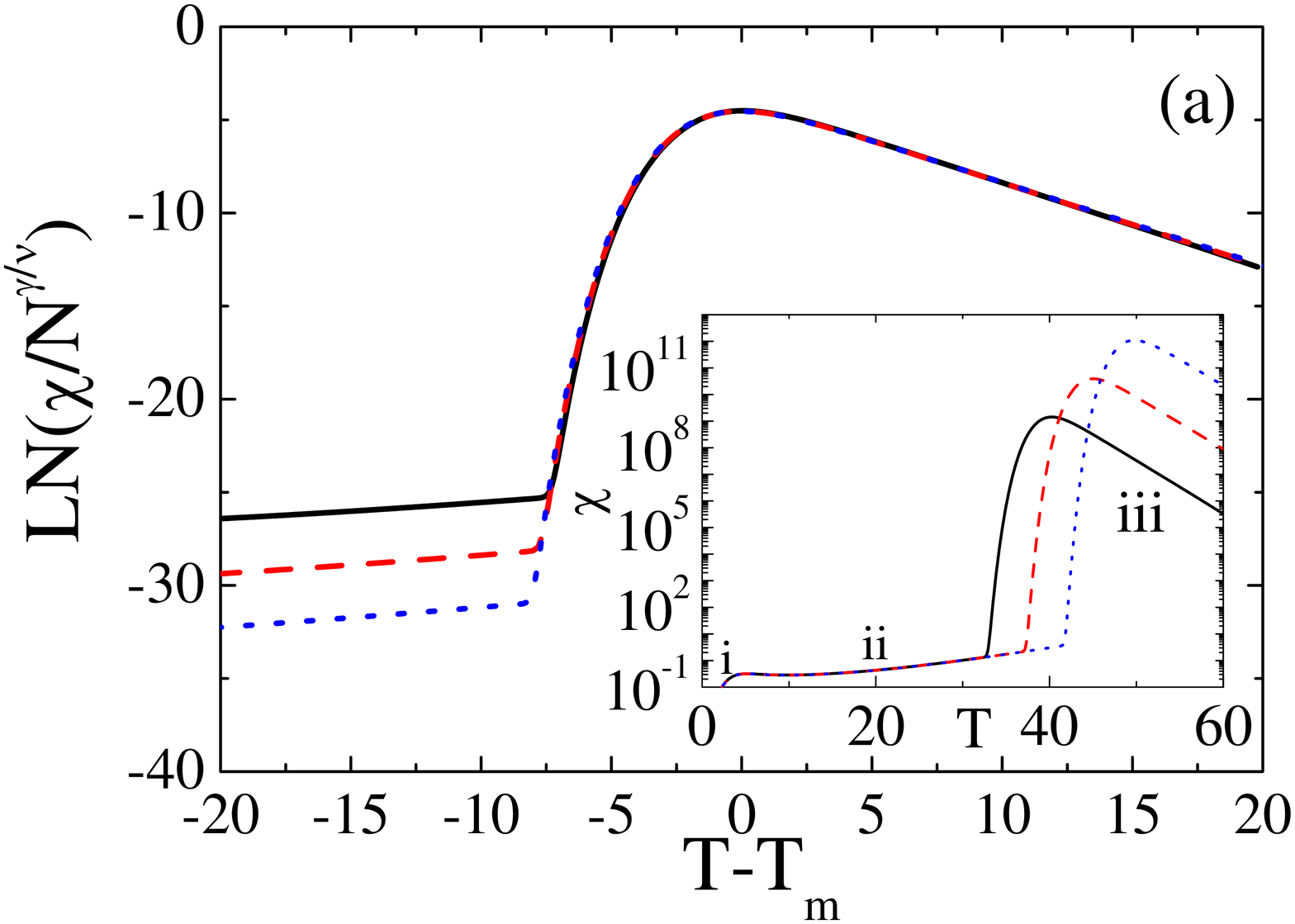}
\includegraphics*[width=8cm,height=5.5cm,angle=0]{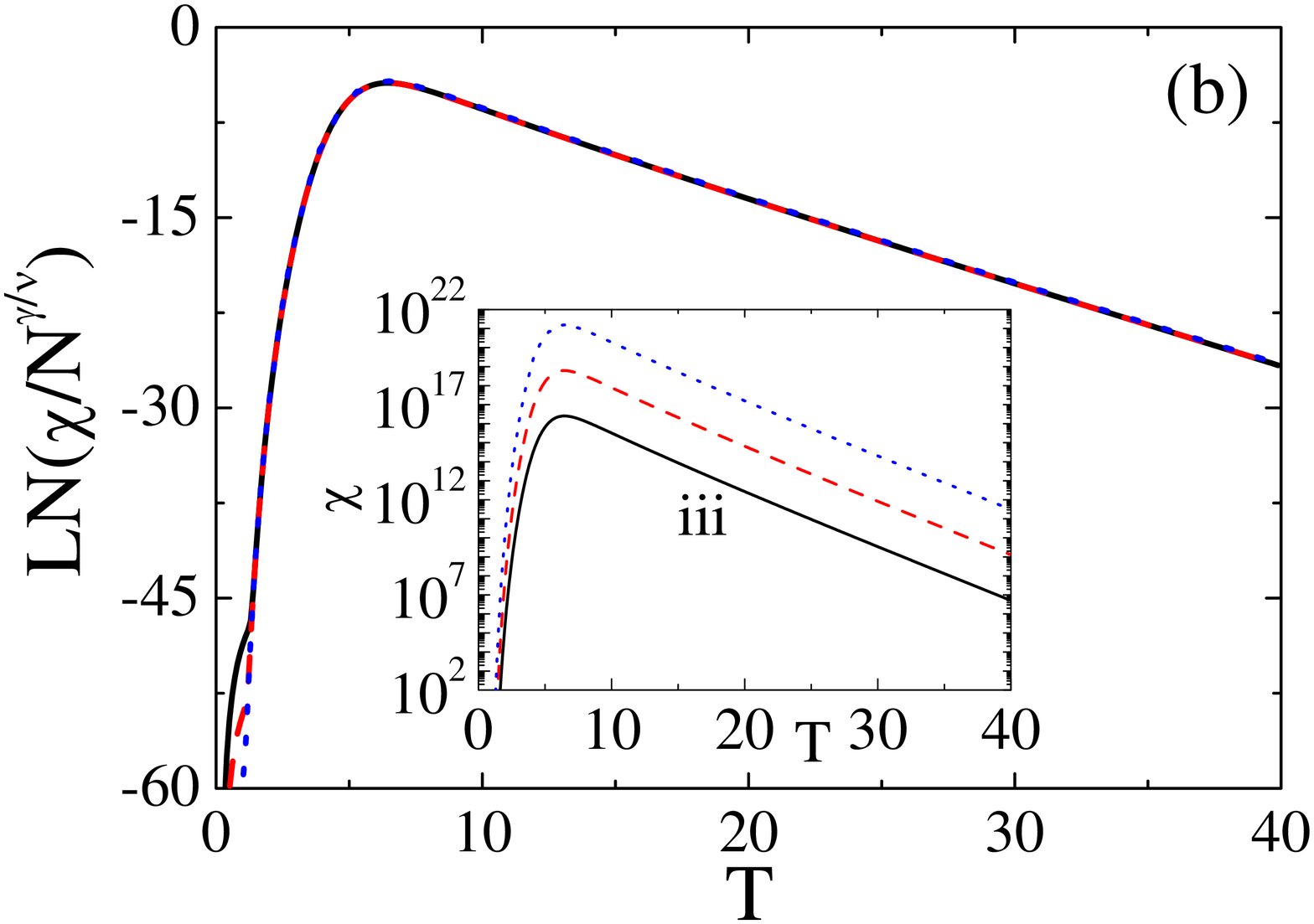}
\includegraphics*[width=8cm,height=5.5cm,angle=0]{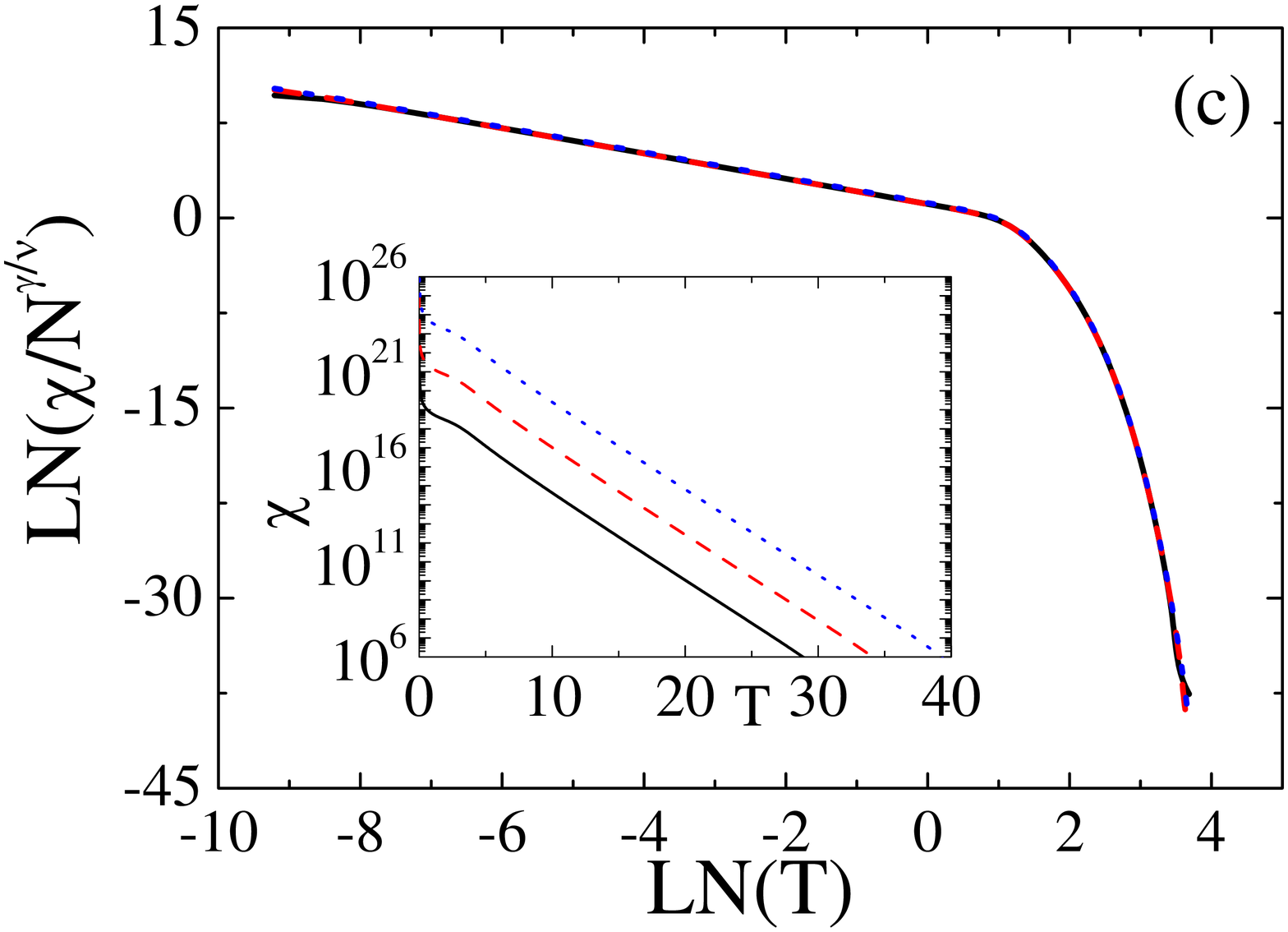}
\end{center}
\caption{Main panels show scaling properties of $\chi(T,h=0;N(g))$ against
$T$ for different values of $g=40$ (solid), 45(dashes), and 50(dots), when
$\mu=0.5$ (a), $1.0$ (b), and $1.5$ (c). The inserts show the same curves
in the original variables. (a) Scaling properties of $\chi$ as function of
$T-T_m$, with exponent $\gamma/\nu=0.607$, are observed in the region iii
of the magnetization $m$ (see Fig. 5a). (b) Scaling properties of $\chi$
as function of $T$, with exponent $\gamma/\nu=1.0$, valid in the region
iii, with exponential decrease of $m$, as discussed in Fig. 5b. (c)
Scaling properties of $\chi$ as function of $T$, with exponent
$\gamma/\nu=1.0$. Scaling is valid in the two regions shown in Fig. 5c.
The horizontal axis is in logarithmic scale to evidence Curie's law with
$g$ dependent constant $\mathcal{C}_g$. }\label{fig6}
\end{figure}

In the last range of parameter values, $T_d$ and $T_m$ decrease with
respect to $g$. As shown by Fig. 4c, both values converge exponentially
to $0$ with respect to $g$. Fig. 4c also shows that the rate in the
exponential increases with $\mu$.

Therefore, the behavior of $1/ln(\lambda_1/\lambda_2)$ is different from
that one observed for $\mu=1$, the divergence when $T\rightarrow 0$
becoming slower at increasing values of $g$. This suggests that, when
$g\rightarrow\infty$, any collective spin ordering is weaker than that of
an Ising chain, rather typical for a paramagnetic situation.

The shape of the $m$ curves becomes completely different. The stable
plateau at $m=1$ for a finite temperature interval,  which survived until
$\mu=1$, disappears as $g$ increases, indicating that no spontaneous
magnetization exists for a $T>0$. This suggests that, in the thermodynamic limit,
$m\equiv0\: \forall\, T>0$. Our results do not allow to assert that
$m\equiv0$ also for $T=0$.

The behavior of $\chi$ supports the conclusion of a low temperature
paramagnetic phase. It increases very rapidly from 0 to a maximum value at
$T_m$, followed by a $1/T$ law. Since $T_m$ goes exponentially to zero, a
Curie law prevails for large $g$. This is shown by the scaled curves in
Fig. 6c, which indicate that the Curie constant $\mathcal{C}$ depends on
$g$.

\section {Discussion and Conclusions}

The results we obtained for the magnetic behavior of the Ising model with
node dependent interaction constants reveal a quite rich picture, although
no critical behavior at a finite temperature has been identified. The
properties of specific heat show that the $g$ dependent curves converge
very rapidly to a well defined value in the thermodynamic limit. On the
other hand, magnetization and susceptibility indicate a much more complex
behavior which, for certain temperature intervals, are heavily dependent
on the value of $g$.

The TM method allows for the comparison of $m$ and $\chi$ for
different values of $g$, which leads to the identification that part of
the results are due to finite size events. The curves showing such
effects are amenable to very precise collapsing by adequate scaling
expressions, similar to critical
points in magnetic models on Euclidian lattices. This includes the
dependence of characteristic values of the temperature ($T_d$ and $T_m$)

The behavior of the system in the region $\mu<1$ is close to that observed
for magnetic system with uniform interactions on BA networks: only an
ordered phase is observed at any value of $T$. $\mu=1$ characterizes a
crossover in the behavior of the system, as for $\mu<1$ the magnetization
vanishes, for any value of $T$, when $g\rightarrow \infty$. This region
reveals a typical behavior of a genuine paramagnetic system. This
pictures is corroborated by the behavior of $\chi$, as one finds that a
Curie law is valid in a limited region close to $T=0$. For larger values
of $T$, the decay is characterized by an exponential decay.

In the context of complex networks our most important finding is
that the relation of Eq.(1) between effective topology and
interaction strength proposed in Refs. [7,8] does not have general
validity for all scale-free networks since the Apollonian case
behaves differently.

\section{Acknowledgement}

R.F.S. Andrade and J.S. Andrade Jr. thank CNPq for financial support.

\section{Appendix}

As discussed in the Section II, for any generation $g$, the node occupying
the central position of the AN has the largest degree $k_c(g)$, the value
of which results form the difference equation relating the values of $k_c$
at two successive generations: $k_c(g+1)=2k_c(g)$. The degree of the nodes
at the external corners obey a similar equation, namely:
$k_e(g+1)=2k_e(g)-1$.

The AN can be disassembled in triangles, in such a way that each node $i$
of degree $k_i$ belongs to $k_i$ triangles. The only exception refers to
the nodes at the external corners, which have degree $k_e(g)$ but belong
to $k_e(g)-1$ triangles. Each such triangle can be characterized by the
degree of its three nodes. The number $P(g)$ of different triangles at
generation $g$ can be expressed in terms of $\rho(g)$ and $\tau(g)$,
respectively the number of triangles that does not include (includes) an
external node: $P(g)=\rho(g) + \tau(g)$. Since they obey the relations
$\rho(g+1)=\rho(g)+\tau(g)-1$ and $\tau(g+1)=\tau(g)+1$, we obtain
$\rho(g)=(g^2-3g+2)/2$ and $\tau(g)=g$, from which the expression for
$P(g)$ anticipated in Section III follows. The number $\rho(g)$ can be
further decomposed in terms of $\iota(g)$ and $\kappa(g)$, respectively
the number of different triangles that includes (does not include) the
central node at generation $g$. It is a simple matter of inspection to see
that $\iota(g)=g-2$ and $\kappa(g)=(g^2-5g+6)/2$, $\forall g\geq 2$.

For the purpose of computing the TM's, it is necessary to identify the
distinct triangles present in the AN. This proceeds by the collection
$T_{g,\zeta}(k_1,k_2,k_3)$, where $g$ indicates the generation, $\zeta$ is
a number $\in[1,P(g)]$, and $k_i$ indicate the degrees of the nodes at the
vertices of the triangle. $T_{g,\zeta}(k_1,k_2,k_3)$ are recursively
defined according to the following rules:

1) $k_3=3,\,\forall g, \,\forall \zeta$.

2) Since $\tau(g)=g$, one single new triangle containing an external node
is introduced into the network, $\forall g$. We use $\zeta=P(g-1)+1$ to
characterize it, and note further that $k_1=k_e(g),\,k_2=k_c(g)$. Any such
triangle retains the $k_i$ values for all further generations, i.e.:
\begin{equation}\label{eqa1}
T_{g,\zeta}(k_1,k_2,k_3)=T_{g,\zeta '}(k_1,k_2,k_3),\,\forall \zeta ' \geq
\zeta.
\end{equation}

3) $\forall g$, there are $\rho(g)$ different triangles, among which
$\kappa(g)$ have been introduced in previous generations. They will be
characterized by the same values of $k_i$, so that Eq. (\ref{eqa1}) also
holds for this subset. The remaining $\iota(g)$ new triangles are numbered
according to the rule: $\zeta=P(g-1)+\ell,\,\ell=2,3,...,g-1$. For each
value of $\ell$, we set $k_1=k_c(g),\,k_2=k_c(\ell)$.

The final step consists in establishing the rule to combine the
contributions to the partition function from three distinct triangles at
generation $g$ to obtain the partition function at generation $g+1$
according to Eq. (\ref{eq6}). If the properties of the systems are to be
computed until a chosen value $g$, we are required to start with $P(g)$
distinct triangles, precisely identified as discussed above. Then, as
discussed in Section III, it is necessary to define a map that selects the
proper values of $\eta,\,\epsilon,\,\delta \in [1,P(g)]$ used to perform
the trace over the common central node of the triangle $\alpha \in
[1,P(g-1)].$ So let us note that
\begin{equation}\label{eqa2}
[1,P(g)]=[1] \bigcup \left \{ \bigcup _{j=2}^g
[P(g-1)+1,P(g-1)+j-1]\right\}.
\end{equation}

Then, the values of $\eta,\,\epsilon,\,\delta $ are given, as function of
$\alpha$, by the following expressions:

\begin{equation}\label{eqa3}
\begin{array}{l}
\nonumber   \alpha = 1:\\
\nonumber  \eta = 1,\, \epsilon=\delta=2; \\
\\
\nonumber  \alpha = P(j-1)+1, j\in [2,g-1]:\\
\nonumber  \eta = 2,  \epsilon= P(j)+1, \delta=P(j)+2; \\
\\
\nonumber  \alpha \in [P(j-1)+2,P(j-1)+j-1], j\in [2,g-1]:\\
\nonumber  \eta = P(\alpha-P(j-1))+2,  \epsilon= P(j)+2, \delta=\alpha +
j.
\end{array}
\\
\end{equation}

As the AN is self similar, these maps are also valid for all forthcoming
partial trace operations, until only one single triangle is left. At this
step, the remaining TM contains the contribution form all spin
configurations, from which the thermodynamical properties follow.

With the help of these relations, a set of recurrence maps can derived
from Eq. (\ref{eq5}), which allow for the evaluation of the free energy
and its derivatives:

\begin{equation}
\label{eqa4}
\begin{array}{l}
f_{g+1}^\alpha=\frac{N_g}{N_{g+1}}(f_g^\eta+f_g^\epsilon+f_g^\delta) -
\\
\frac{T}{N_{g+1}}\{\ln(1+ x_{1,g}^\eta x_{1,g}^\epsilon x_{1,g}^\delta
\exp(-2 \beta
h))\}\\
- \frac{h}{N_{g+1}};
\end{array}
\end{equation}

\begin{equation}
\label{eqa5} x_{1,g+1}^\alpha=\frac{x_{2,g}^\epsilon x_{2,g}^\delta +
x_{1,g}^\eta x_{3,g}^\epsilon x_{3,g}^\delta \exp(-2 \beta h))}
{1+x_{1,g}^\eta x_{1,g}^\epsilon x_{1,g}^\delta \exp(-2 \beta h));}
\end{equation}

\begin{equation}
\label{eqa6} x_{2,g+1}^\alpha=\frac{x_{2,g}^\eta x_{4,g}^\delta +
x_{3,g}^\eta x_{1,g}^\epsilon x_{2,g}^\delta \exp(-2 \beta h))}
{1+x_{1,g}^\eta x_{1,g}^\epsilon x_{1,g}^\delta \exp(-2 \beta h));}
\end{equation}

\begin{equation}
\label{eqa7} x_{3,g+1}^\alpha=\frac{x_{2,g}^\eta x_{2,g}^\epsilon
x_{6,g}^\delta + x_{3,g}^\eta x_{3,g}^\epsilon x_{7,g}^\delta \exp(-2\beta
h))} {1+x_{1,g}^\eta x_{1,g}^\epsilon x_{1,g}^\delta \exp(-2 \beta h))}
\end{equation}

\begin{equation}
\label{eqa8} x_{4,g+1}^\alpha=\frac{x_{4,g}^\eta x_{4,g}^\epsilon +
x_{5,g}^\eta x_{5,g}^\epsilon x_{1,g}^\delta \exp(-2 \beta h))}
{1+x_{1,g}^\eta x_{1,g}^\epsilon x_{1,g}^\delta \exp(-2 \beta h));}
\end{equation}

\begin{equation}
\label{eqa9} x_{5,g+1}^\alpha=\frac{x_{4,g}^\eta x_{6,g}^\epsilon
x_{2,g}^\delta + x_{5,g}^\eta x_{7,g}^\epsilon x_{3,g}^\delta \exp(-2\beta
h))} {1+x_{1,g}^\eta x_{1,g}^\epsilon x_{1,g}^\delta \exp(-2 \beta h));}
\end{equation}

\begin{equation}
\label{eqa10} x_{6,g+1}^\alpha=\frac{x_{6,g}^\eta x_{4,g}^\epsilon
x_{4,g}^\delta + x_{7,g}^\eta x_{5,g}^\epsilon x_{5,g}^\delta \exp(-2\beta
h))} {1+x_{1,g}^\eta x_{1,g}^\epsilon x_{1,g}^\delta \exp(-2 \beta h));}
\end{equation}

\begin{equation}
\label{eqa11} x_{7,g+1}^\alpha=\frac{x_{6,g}^\eta x_{6,g}^\epsilon
x_{6,g}^\delta + x_{7,g}^\eta x_{7,g}^\epsilon x_{7,g}^\delta \exp(-2\beta
h))} {1+x_{1,g}^\eta x_{1,g}^\epsilon x_{1,g}^\delta \exp(-2 \beta h)).}
\end{equation}

In the above relations, the following variables have been used:

\begin{equation}
\label{eqa12}
x_{1,g}=\frac{q_g}{p_g};\:x_{2,g}=\frac{r_g}{p_g};\:x_{3,g}=\frac{s_g}{p_g};\:
\end{equation}

\begin{equation}
\label{eqa13}
x_{4,g}=\frac{t_g}{p_g};\:x_{5,g}=\frac{u_g}{p_g};\:x_{6,g}=\frac{v_g}{p_g};\:
\end{equation}

\begin{equation}
\label{eqa14} x_{7,g}=\frac{w_g}{p_g}.
\end{equation}

\bibliographystyle{prsty}

\end{document}